\documentclass[a4paper,12pt]{article}
\usepackage{amsbsy}
\usepackage[fleqn]{amsmath}
\usepackage{amssymb}
\usepackage{array}
\usepackage{booktabs}
\usepackage{xcolor}
\usepackage[colorlinks, linkcolor=blue, anchorcolor=blue, citecolor=blue]{hyperref}
\usepackage{mathrsfs}
\usepackage{tabularx}
\usepackage{latexsym}
\usepackage{textcomp}
\usepackage{pstricks}
\usepackage{colortbl}
\usepackage{multirow}
\usepackage{fancyhdr}
\usepackage{bm}
\usepackage{booktabs}
\usepackage{subfigure}
\usepackage{caption}
\usepackage{soul}
\usepackage{cleveref}
\usepackage{cite}
\usepackage{lineno}
\usepackage{ulem}
\usepackage{graphicx}
\usepackage{epstopdf}
\usepackage{indentfirst}
\usepackage{setspace}
\usepackage[latin1]{inputenc}
\usepackage{titlesec}
\usepackage{tabularx} 
\usepackage{ragged2e} 
\usepackage{booktabs} 
\usepackage{float}  
\usepackage{stmaryrd}
\usepackage{amsthm}
\usepackage{amssymb}
\SetSymbolFont{stmry}{bold}{U}{stmry}{m}{n}

\titleformat*{\section}{\large\bf}
\titleformat*{\subsection}{\normalsize\bf}
\crefname{section}{Section}{Sections}
\crefname{equation}{Eq.}{Eqs.}
\crefname{figure}{Figure}{Figs.}
\crefname{table}{Table}{Tables}
\newcommand{\RomanOne}{\mathbf{\uppercase\expandafter{\romannumeral1}}}
\newcommand{\RomanTwo}{\mathbf{\uppercase\expandafter{\romannumeral2}}}

\begin{document}

\newtheorem{theorem}{Theorem}[section]
\theoremstyle{definition}
\newtheorem{remark}[theorem]{Remark}

\hypersetup{
    colorlinks=true,
    linkcolor=blue,
    urlcolor=blue,
    citecolor=blue
}

%
%
\begin{center}
\Large\textbf{LENNs: Locally Enhanced Neural Networks for High-Fidelity Modeling in Solid Mechanics}
\end{center}
\large{
\begin{center}
\textbf{Zhihong Lai}$^{a}$\footnotemark[1], \textbf{Luyang Zhao}$^{a}$\footnotemark[1], \textbf{Qian Shao}$^{a,b}$\footnotemark[2]
\end{center}
}
\small{
\begin{center}
$^a$Department of Engineering Mechanics, School of Civil Engineering, Wuhan University, 430072 Wuhan, China\\
$^b$Wuhan University Shenzhen Research Institute, 518057 Shenzhen, China\\
\end{center}
}

\footnotetext[1]{These authors contributed equally to this work.}
\footnotetext[2]{Corresponding author. E-mail address: qian.shao@whu.edu.cn.}
%
%
\begin{flushleft}
\large\textbf{Abstract}
\end{flushleft}
\indent\indent 
Despite prior advances in PINNs, significant challenges remain in  localized solid mechanics problems because of the limitations of single network formulations in simultaneous resolution of smooth global responses and near-tip singularities, and inadequacy in discontinuity representation, leading to unstable training and limited accuracy. To address the challenges, we propose Locally Enhanced Neural Networks (LENNs) that characterize localized discontinuities in solid mechanics via multilevel modeling. In particular, this novel
framework employs a global network for the bulk solution and activates a local network in localized area for non-smooth response, coupled through a smooth window function that enables weighted superposition of local and global solutions. Moreover, the local network embeds additional functions that encode the discontinuous information into the input to capture localized non-smooth mechanical behaviors. Finally, the composite solution is substituted into the total potential energy functional for unified optimization. With this structure, the method resolves the conflict of single network in representing both smooth global and singular local fields without additional interface-loss terms and amplifies the contribution of localized critical features in energy optimization. We focus on a series of numerical experiments in solid mechanics to demonstrate the performance of the method. Results show that LENNs perform well in addressing localized discontinuous problems and provide accurate predictions for both displacement and stress fields.

\begin{flushleft}
\end{flushleft}

%
%
\section{Introduction}\label{introduction}

In solid mechanics, localized features induced by discontinuities, such as singular stress fields near crack tips, displacement jumps across crack surfaces, and stress jumps across material interfaces, are of critical importance in various engineering applications, as they often govern the primary sites of stress concentration or failure initiation. A key characteristic of these problems is that while local effects may have a limited influence on the overall structural response, they significantly affect the mechanical behavior in their immediate vicinity. Accurately capturing these localized phenomena is essential in solid mechanics and requires high-resolution modeling techniques and flexible numerical representations. Conventional numerical methods, particularly mesh-based methods \cite{zienkiewicz2005finite}, face significant difficulties in modeling highly localized behaviors. Firstly, simulating evolving discontinuities, such as crack propagation, is challenging as it requires continuous remeshing to conform to the moving interface, which is computationally expensive and prone to inaccuracies. For even static discontinuities, achieving sufficient accuracy typically requires highly refined meshes and often the incorporation of specialized elements. Furthermore, in parametric studies involving variations in materials properties or boundary conditions, these approaches usually necessitate a complete reformulation of the numerical model, which reduces computational flexibility and efficiency, while significantly increasing the pre-processing burden.


Recent advances in deep learning, particularly the development of Physics-Informed Neural Networks (PINNs) \cite{raissi2019physics}, have introduced a promising alternative to overcome the aforementioned limitations. Owing to their mesh-free nature, PINNs eliminate the dependence on fine spatial discretizations, making them naturally suited for problems with irregular geometries. By embedding the governing physical laws directly into the loss function, PINNs provide a unified framework for solving broad classes of partial differential equations (PDEs), largely circumventing the need for problem-specific numerical formulations. The strong nonlinear function approximation capability of neural networks \cite{hornik1989multilayer} further enables them to capture complex solution behaviors that are challenging for traditional methods. Moreover, PINNs offer exceptional flexibility for scenarios involving varying parameters, such as changes in boundary conditions or material properties. This adaptability is primarily facilitated through transfer learning, where a pretrained network can be efficiently fine-tuned to new settings, avoiding computationally expensive full retraining and significantly reducing the computational overhead \cite{haghighat2021physics,goswami2020transfer,wang2025transfer}. PINNs can be formulated using either the strong form \cite{raissi2019physics} or the energy form \cite{yu2018deep} of the PDEs, providing versatile approaches for a wide range of scientific and engineering applications. They have demonstrated considerable advancements in fluid dynamics \cite{mao2020physics,cai2021physics,jin2021nsfnets}, heat transfer \cite{cai2021physics1,laubscher2021simulation}, material design \cite{chen2020physics,liu2019multi}, and are increasingly applied to solid mechanics problems, including linear elasticity \cite{haghighat2021physics, abueidda2021meshless,abueidda2022deep,li2021physics,zhuang2021deep,rao2021physics}, elasto-plasticity \cite{niu2023modeling,he2023deep}, hyperelasticity \cite{fuhg2022mixed,nguyen2021parametric},and fracture mechanics\cite{goswami2020transfer,goswami2022physics,ning2023physics,yu2024nonlocal}.

Despite this progress, the application of PINNs to solid mechanics encounters a fundamental challenge when modeling discontinuities, such as cracks and material interfaces. Unlike intrinsic discontinuities emerging from the PDEs themselves (e.g., shocks in Burgers' equation), cracks and interfaces introduce extrinsic discontinuities through geometric partitioning or material heterogeneity. These manifest as displacement jumps (strong discontinuities) across cracks or derivative jumps (weak discontinuities) across interfaces. Such behavior fundamentally contradicts the continuity assumptions inherent in standard neural network approximation theory \cite{hornik1989multilayer}. Consequently, vanilla PINNs struggle to accurately capture these features, resulting in significant modeling errors and poor convergence.


To address these challenges, several strategies have been proposed, which can be broadly categorized into two groups. The first group employs domain decomposition, dividing the problem domain into non-overlapping subdomains to simplify the learning task. 
Jagtap et al.~\cite{jagtap2020conservative,jagtap2020extended} developed extended PINNs and conservative PINNs based on this idea. Wang et al.~\cite{wang2022cenn} proposed a conservative energy method based on neural networks (CENN) with domain decomposition for modeling mode III cracks and heterogeneities. Gu et al.~\cite{gu2023enriched} enhanced this approach by enriching the solution space near crack tips using analytical asymptotic functions to accurately resolve the singular stress field. Sarma et al.~\cite{sarma2024interface} introduced interface PINNs, which employ distinct activation functions across subdomains to better capture solution behaviors across interfaces in elliptic problems. The second group seeks to enhance the representational capacity of the neural network itself, often within a single-domain setup. Zhao et al.~\cite{zhao2025denns} proposed Discontinuity-Embedded Neural Networks (DENNs), which incorporate features constructed from signed distance functions (SDFs) into the network inputs to explicitly model both strong and weak discontinuities. Similarly, Tseng et al. \cite{tseng2023cusp} introduced an extra input to capture cusp-like discontinuous behaviors. Other innovative approaches include mixed formulations \cite{rezaei2022mixed} that separately approximate displacements and stresses to enhance accuracy, and methods that directly incorporate interface conditions into the loss function \cite{yao2023deep}. Separately, to improve overall training performance, Moseley et al.~\cite{moseley2023finite} employed multiple neural networks over overlapping subdomains, aggregating their outputs to approximate the global solution.

However, a critical challenge persists: when discontinuities are highly localized, occupying a small fraction of the global domain (e.g., a short crack with a characteristic length less than 1/10 of the domain size, or a sharp crack tip with highly concentrated stress field), existing PINNs-based methods often fail to deliver satisfactory accuracy. For instance, Zhao et al.~\cite{zhao2025denns} reported that as the ratio of crack length to domain size decreases to approximately 1/10, the accuracy of DENNs in capturing the singular crack-tip stress field deteriorates markedly. This failure stems from several factors. First, a single network architecture struggles to simultaneously characterize the smooth global response and the highly nonlinear or singular local behavior, resulting in a trade-off between local accuracy or global consistency. Typically, resolving the localized region influenced by the discontinuity requires a disproportionately dense distribution of dense collocation points and larger networks. Second, when discontinuities are spatially confined, their contribution to the global energy norm is minimal. Consequently, these critical local features are easily overwhelmed during optimization, leading to under-fitting in these regions. Third, neural networks exhibit a pronounced spectral bias \cite{wang2021eigenvector}, favoring the learning of low-frequency components first, which impedes the accurate resolution of the high-frequency features associated with localized stress concentrations and singularities. While non-overlapping domain decomposition methods may alleviate the issue of energy imbalance by isolating the discontinuity, they face their own set of challenges. The requirement to impose interface conditions between subdomains introduces additional constraints into an already complex multi-objective optimization problem, often complicating convergence of neural networks. Furthermore, although general techniques like Fourier feature mappings \cite{wang2021eigenvector} or overlapping domain decompositions \cite{moseley2023finite} have been proposed to mitigate spectral bias and improve training, their direct application to problems with strong geometric discontinuities is non-trivial. This is because their inherent continuity-enforcing mechanisms are fundamentally incompatible with the presence of a crack-induced geometric discontinuity.

To address these limitations, we propose a Locally Enhanced Neural Networks (LENNs) framework specifically engineered to capture highly localized discontinuities. Our approach employs two distinct neural networks, including a global network that captures the bulk solution and a local network dedicated to resolving the solution within a user-prescribed region surrounding the discontinuity. Specifically, the input to the local network is augmented with discontinuous embeddings constructed from signed distance functions to explicitly encode the discontinuous geometry, enabling the neural network to produce solutions with expected local behaviors. These two networks are coupled through a smooth spatial window function. The output of the local network is multiplied by this window function and additively superimposed onto the output of the global network to form the composite solution. Consequently, the local network is selectively activated only within the region of interest. To enhance numerical stability, the spatial coordinates input to each network are independently normalized to $[-1, 1]$ based on their respective domains. The governing physics are enforced using the energy formulation of PDEs, where the composition solution is directly substituted into the total potential energy functional of the system to define the loss function for unified optimization. This approach automatically satisfies natural boundary conditions. The proposed LENNs provide several key advantages. The dual-network decomposition resolves the functional conflict. The global network is free to learn the smooth, low-frequency bulk response, while the local network specializes in the high-frequency, singular behavior near the discontinuity. The window function ensures a smooth and stable functional decoupling. The local network is confined to its region, preventing it from degrading the far-field solution predicted by the global network, and vice versa. During training, the total energy functional is computed over the entire domain. A key design feature is that the spatial window function confines the output of the local network to its designated region. Consequently, the gradients of the loss function with respect to the parameters of the local network are dominated by the energy contribution from within this region. This architectural choice, combined with the fact that the global network efficiently minimizes the energy in the far-field, means that the optimization process effectively amplifies the relative weight of the discontinuity-induced energy. This mechanism prevents the local critical features from being overlooked during training, thereby mitigating under-fitting. Moreover, decomposing the solution allows the global and local networks to specialize in different frequency bands, i.e., the global network in low-frequency bulk response and the local network in high-frequency localized features, which can alleviate the spectral bias inherent in single-network models. By assigning the learning of high-frequency local features to a dedicated network with locally normalized coordinates, LENNs facilitate a more effective and rapid convergence on the challenging components of the solution, which single-network models struggle to capture. Notably, the framework offers a natural pathway for simulating evolving discontinuities, as the crack path could be advanced by calculating stress intensity factors from the solved displacement field, followed by an iterative update of the window function and discontinuity embeddings that define the new crack geometry.

The paper is organized as follows. \cref{Preliminaries} outlines the governing equations for elastic problems and introduced the fundamental concepts of PINNs. In \cref{Methodology}, we present the proposed LENNs framework. \cref{Results} demonstrates the efficacy of LENNs through several numerical examples in solid mechanics. Finally, \cref{conclusion} summarizes the main findings, discusses the implications of this work, and suggests potential directions for future research.

%
%
\section{Theoretical fundamentals}\label{Preliminaries}

\subsection{Governing equations for linear elasticity with localized features}\label{governing}

In solid mechanics, a primary objective is to determine the mechanical response of a body under prescribed loads. For a continuum occupying a domain $\Omega$, the governing equations of linear elasticity comprise the equilibrium equation, the kinematic relation, and the constitutive law. The equilibrium state is described by:
\begin{equation}
\begin{aligned}
    &\nabla \cdot \boldsymbol{\sigma} + \boldsymbol{f} = \mathbf{0} 
    && \text{in } \Omega, \\[6pt]
    &\boldsymbol{u} = \overline{\boldsymbol{u}} 
    && \text{on } \Gamma_u \quad \text{(Essential B.C.)}, \\[6pt]
    &\boldsymbol{\sigma} \cdot \boldsymbol{n} = \bar{\boldsymbol{t}} 
    && \text{on } \Gamma_t \quad \text{(Natural B.C.)},
\end{aligned}
\end{equation}
where $\boldsymbol{\sigma}$ is the Cauchy stress tensor, $\boldsymbol{f}$ is the body force vector, $\boldsymbol{u}$ is the displacement vector, $\overline{\boldsymbol{u}}$ is the prescribed displacement on $\Gamma_u$, $\bar{\boldsymbol{t}}$ is the prescribed traction on $\Gamma_t$, and $\boldsymbol{n}$ denotes the outward unit normal vector. The infinitesimal strain tensor $\boldsymbol{\varepsilon}$ is linked to the displacement by the kinematic equation:
\begin{equation}
\boldsymbol{\varepsilon} = \frac{1}{2} \left( \nabla \boldsymbol{u} + \nabla \boldsymbol{u}^\top \right).
\end{equation}
For a linear elastic material, the constitutive behavior is governed by the generalized Hooke's law:
\begin{equation}
\boldsymbol{\sigma} = \boldsymbol{C}: \boldsymbol{\varepsilon},
\end{equation}
where $\boldsymbol{C}$ is the fourth-order elasticity tensor. 

\begin{figure}[!hbtp]
        \centering        \includegraphics[width=0.45\textwidth]{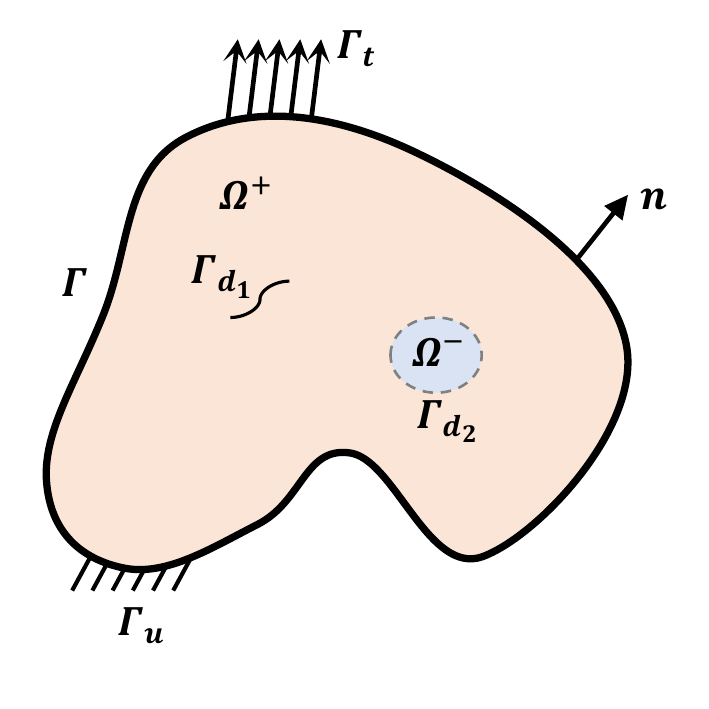}
        \caption{Schematic of a solid domain containing localized features.
        }
        \label{fig1}
\end{figure}

These equations define a well-posed problem for a continuous body subject to small strains and free of internal discontinuities. However, engineering structures frequently involve localized features that introduce discontinuities, such as cracks, inclusions, and material interfaces. While the global response of the structure remains governed by the classical equations, these localized features introduce highly nonlinear and often singular mechanical behaviors within confined regions, posing significant challenges for numerical simulation. This work focuses on two specific types of localized discontinuities, i.e., cracks and material interfaces, which serve as prototypes for strong and weak discontinuities, respectively. As illustrated in \cref{fig1}, these features act as internal boundaries where specialized conditions apply. On a crack surface $\Gamma_{d_1}$, the displacement field exhibits a strong discontinuity. On a material interface $\Gamma_{d_2}$, the displacement continuity is maintained, and the traction must be in equilibrium. Due to the change in material properties across the interface, these conditions typically lead to a jump in the normal derivative of the displacement ($\llbracket \partial\boldsymbol{u} / \partial \boldsymbol{n} \rrbracket \neq \mathbf{0}$). These local effects are incorporated through additional conditions imposed on the internal surfaces:
\begin{equation}
        \begin{aligned}
           &\llbracket \boldsymbol{u} \rrbracket \neq \mathbf{0}\text{ and } \boldsymbol{\sigma} \cdot \boldsymbol{n} = \bar{\boldsymbol{t}}_{d_1} &\text{ on }\Gamma_{d_1}, \\[6pt]
           &\llbracket \boldsymbol{u} \rrbracket = \mathbf{0} \text{ and } \llbracket \boldsymbol{\sigma} \cdot \boldsymbol{n} \rrbracket = \mathbf{0} &\text{ on }\Gamma_{d_2} ,
           \label{interface_condition}
        \end{aligned}
\end{equation} 
where $\llbracket \cdot \rrbracket$ denotes the jump operator, and $\bar{\boldsymbol{t}}_{d_1}$ is the traction on the crack surfaces. Typically, a traction-free condition is imposed on the crack surfaces, i.e., $\boldsymbol{\sigma} \cdot \boldsymbol{n} = \mathbf{0}$ on $\Gamma_{d_1}$. 

The principle of minimum potential energy provides a variational framework well-suited for this class of problems. The total potential energy $\Pi$ of the system, including the localized effects of internal boundaries, is given by:
\begin{equation}
\begin{aligned}
\Pi(\boldsymbol{u}) = \int_{\Omega} \frac{1}{2}\boldsymbol{\varepsilon}(\boldsymbol{u}) : \boldsymbol{C} : \boldsymbol{\varepsilon}(\boldsymbol{u}) d\Omega - \int_{\Omega} \boldsymbol{u} \cdot \boldsymbol{f} d\Omega - \int_{\Gamma_t} \boldsymbol{u} \cdot \bar{\boldsymbol{t}} d\Gamma
- \int_{\Gamma_{d_1}} \left( \boldsymbol{u}^{+} \cdot \bar{\boldsymbol{t}}_{d_1}^+ + \boldsymbol{u}^{-} \cdot \bar{\boldsymbol{t}}_{d_1}^- \right) d\Gamma.
\end{aligned}
\end{equation}
Finding the displacement field $\boldsymbol{u}$ that minimizes this functional is equivalent to solving the strong-form equations. The stationary condition $\delta\Pi = 0$ naturally satisfies the Neumman boundary conditions on $\Gamma_t$ and the traction continuity condition on $\Gamma_{d_2}$ \cite{zhao2025denns}. Critically, for the common case of traction-free cracks ($\bar{\boldsymbol{t}}_{d_1} = \mathbf{0}$), the final integral vanishes, reducing the potential energy to the classical form:
\begin{equation}
\label{eq:weak_form_final}
\Pi(\boldsymbol{u}) = \int_{\Omega} \frac{1}{2}\boldsymbol{\varepsilon}(\boldsymbol{u}) : \boldsymbol{C} : \boldsymbol{\varepsilon}(\boldsymbol{u}) d\Omega - \int_{\Omega} \boldsymbol{u} \cdot \boldsymbol{f} d\Omega - \int_{\Gamma_t} \boldsymbol{u} \cdot \bar{\boldsymbol{t}} d\Gamma.
\end{equation}
Although the explicit contribution of the crack to the functional has disappeared, the true solution $\boldsymbol{u}$ that minimizes this energy must still implicitly capture the displacement jump across $\Gamma_{d_1}$ and the stress singularity at the crack tip. The globally integrated energy functional in \cref{eq:weak_form_final} provides no direct, localized information to guide a solver toward these critical localized features. This lack of explicit guidance is a primary source of failure for methods that rely solely on minimizing this global objective, including energy-based PINNs.

\subsection{Energy-based formulation for Physics-Informed Neural Networks}\label{PINNs}

Physics-Informed Neural Networks (PINNs) provide a mesh-free framework for solving PDEs by encoding the governing physical laws into the loss function of a neural network. In this framework, the unknown solution $\boldsymbol{u}$ is approximated by a deep neural network, denoted as $\boldsymbol{u}^*(\boldsymbol{x};\boldsymbol{\theta})$, where $\boldsymbol{\theta}$ represents the trainable parameters (weights $\boldsymbol{W}$ and biases $\boldsymbol{b}$). A standard $L$-layer feedforward neural network is defined by the layer-wise transformation:
\begin{equation}
\begin{aligned}
\mathcal{NN}_1(\boldsymbol{x}) &= \varsigma(\boldsymbol{W}_1 \boldsymbol{x} + \boldsymbol{b}_1), \\[6pt]
\mathcal{NN}_k(\boldsymbol{x}) &= \varsigma(\boldsymbol{W}_k \mathcal{NN}_{k-1}(\boldsymbol{x}) + \boldsymbol{b}_k), \quad \text{for } k = 2, \ldots, L-1, \\[6pt]
\boldsymbol{u}^*(\boldsymbol{x}; \boldsymbol{\theta}) &= \boldsymbol{W}_L \mathcal{NN}_{L-1}(\boldsymbol{x}) + \boldsymbol{b}_L,
\end{aligned}
\end{equation}
where $\varsigma(\cdot)$ is a nonlinear activation function. The parameters $\boldsymbol{\theta}$ are optimized by minimizing a loss function that penalizes the residuals of the governing equations.

The specific form of the loss function depends on the PDEs representation. While the strong-form collocation approach is common, the variational principle described in \cref{governing} provides a robust foundation for the PINNs loss function in solid mechanics. We adopt this energy-based formulation (\cref{eq:weak_form_final}), defining the loss $\mathcal{L}(\boldsymbol{\theta})$ as the total potential energy evaluated for the approximation of PINNs:
\begin{equation}
\mathcal{L}(\boldsymbol{\theta}) = \Pi(\boldsymbol{u}^*)
= \int_{\Omega} \frac{1}{2}\boldsymbol{\varepsilon}(\boldsymbol{u}^*) : \boldsymbol{C} : \boldsymbol{\varepsilon}(\boldsymbol{u}^*) d\Omega
- \int_{\Omega} \boldsymbol{u}^* \cdot \boldsymbol{f} d\Omega
- \int_{\Gamma_t} \boldsymbol{u}^* \cdot \overline{\boldsymbol{t}} d\Gamma.\label{eq:potential_energy_loss}
\end{equation}
Training the neural network to minimize $\mathcal{L}(\boldsymbol{\theta})$ seeks the solution that satisfies the principle of minimum potential energy. This energy-based approach offers two significant advantages. First, it requires lower-order derivatives compared to the strong form, which can improve numerical stability and reduce computational cost during backpropagation. Second, Neumann boundary conditions are naturally incorporated as boundary integral terms in the energy functional, eliminating the need for additional soft constraint loss terms and simplifying the optimization process. Essential boundary conditions must still be enforced explicitly, typically by constructing the network output $\boldsymbol{u}^*$ to satisfy them exactly via a smooth extension function.

Despite its elegance, the standard energy-based PINNs approach faces a critical challenge when modeling systems with highly localized discontinuities. As the loss in \cref{eq:potential_energy_loss} is an integral over the entire domain $\Omega$, the energy contribution from a small region containing a discontinuity (e.g., a short crack) can become negligible compared to the total energy of the bulk domain. Consequently, the gradient-based optimization process is insensitive to local errors in these critical regions, leading to the energy imbalance and under-fitting phenomenon discussed in \cref{introduction}. This fundamental limitation motivates the development of our LENNs framework.

\section{Locally Enhanced Neural Networks}\label{Methodology}

\subsection{Dual-network architecture and coupling strategy}\label{LENN1}

\begin{figure}[!hbtp]
    \centering
    \includegraphics[width=0.95\textwidth]{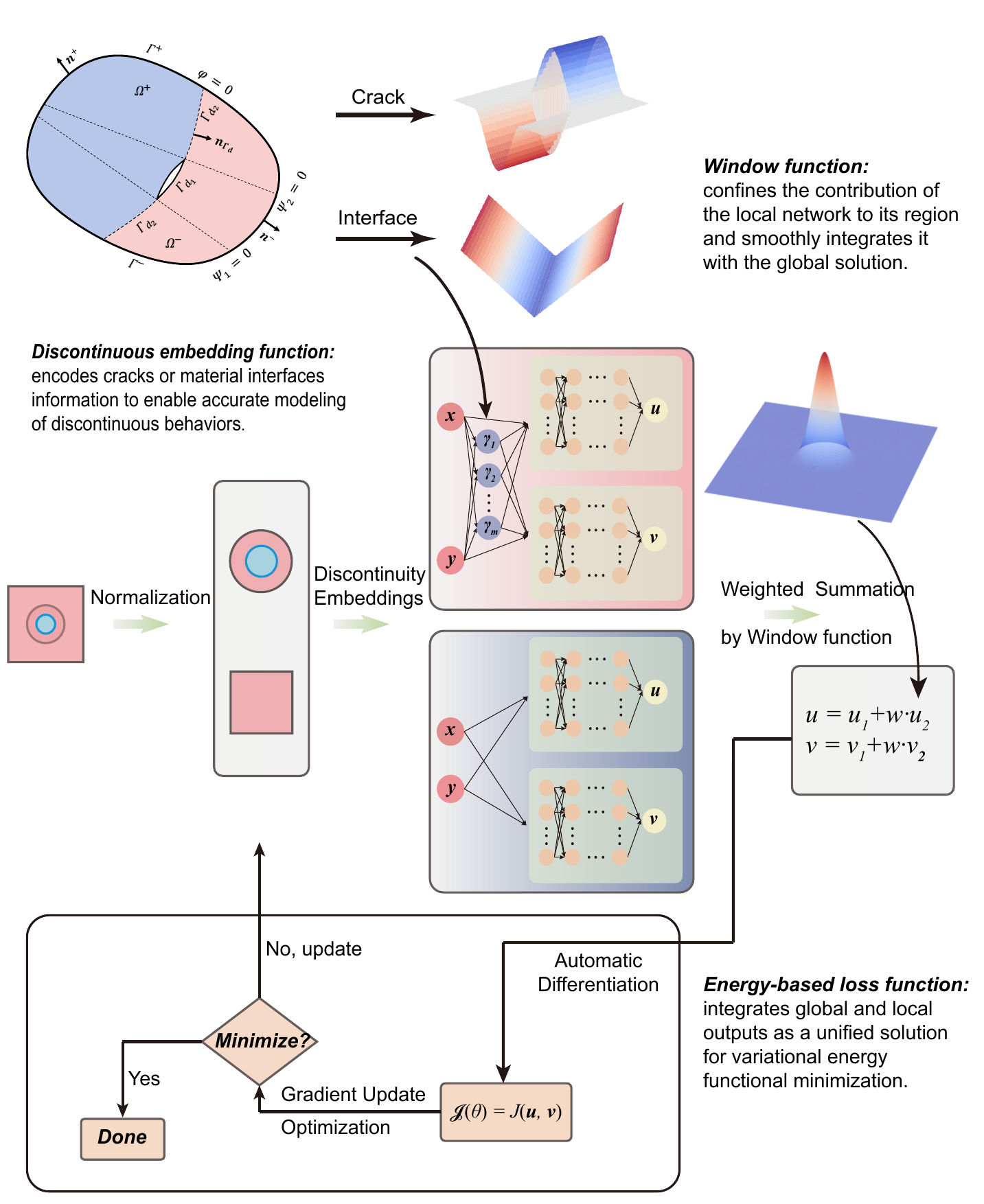}
    \caption{Workflow of LENN, showing the computational path from input coordinates to predicted outputs through the neural architecture.
    }
    \label{LENN}
\end{figure}

As established in \cref{PINNs}, standard energy-based PINNs face a fundamental challenge in modeling highly localized discontinuities due to the energy imbalance between the vast bulk domain and critical local features, leading to under-fitting and an inability to resolve jumps or singularities. To address this, we propose the Locally Enhanced Neural Networks framework. The key insight is to decouple the learning task. A primary network captures the globally smooth solution field, while secondary networks, activated only within user-prescribed regions of interest, are dedicated to resolving the local discontinuities behaviors. This strategy is physically motivated, as critical regions like crack surfaces are often known a priori. The overall workflow is illustrated in \cref{LENN}.

The LENNs framework employs two functionally complementary neural networks. The global network $\mathcal{NN}_\text{global}$ is a relatively compact network that approximates the smooth, background solution field over the entire domain $\Omega$, anaglous to a coarse-scale approximation. The local networks $\mathcal{NN}^i_\text{local}$ are a set of separate, more expressive networks assigned to each critical region. To explicitly encode geometric discontinuities, the input to each local network is augmented with its corresponding embeddings $\gamma^i(\boldsymbol{x})$ constructed from signed distance functions. Thus, the output of the $i$-th local network is given by:
\begin{equation}
\boldsymbol{u}^i_{\text{local}} = \mathcal{NN}^i_{\text{local}}(\boldsymbol{x}, \gamma^i(\boldsymbol{x}); \boldsymbol{\theta}^i_{\text{local}}).
\end{equation}
The input augmentation provides the network with direct geometric knowledge of the discontinuity, enabling it to learn the associated strong or weak discontinuity effectively.

The global and local networks are coupled via an additive superposition scheme, mediated by smooth spatial window functions. This design ensures that the influence of local network is confined to its designated region, reducing redundant model complexity in smooth regions and enhancing approximation fidelity in areas with localized discontinuities. The complete solution approximation is:
\begin{equation}
\boldsymbol{u}(\boldsymbol{x}) = w_\text{global}(\boldsymbol{x}) \cdot \mathcal{NN}_\text{global}(\boldsymbol{x}; \boldsymbol{\theta}_\text{global}) + \sum_{i=1}^{n} w^i_\text{local}(\boldsymbol{x}) \cdot \mathcal{NN}^i_\text{local}(\boldsymbol{x}, \gamma^i(\boldsymbol{x}); \boldsymbol{\theta}^i_\text{local}),
\label{eq:total_solution}
\end{equation}
where $w^i_\text{local}(\boldsymbol{x})$ is the window function for the $i$-th local network. Since the global network operates over the entire domain, its effective window function is unity, $w_\text{global}(\boldsymbol{x}) \equiv 1$. In this work, local regions are defined as circular subdomains for simplicity. The window function  $w^i_\text{local}(\boldsymbol{x})$ must be at least $C^2$-continuous to avoid introducing artificial discontinuities into the stress field. Thus, we employ a cubic B-spline function:
\begin{equation}
w(z) =
\begin{cases}
\frac{2}{3} - 4z^2 + 4z^3, & 0 \le z \le \frac{1}{2} \\
\frac{4}{3} - 4z + 4z^2 - \frac{4}{3}z^3, & \frac{1}{2} < z \le 1, \\
0, & z > 1
\end{cases}
\label{eq:window_function}
\end{equation}
where $z = \| \boldsymbol{x}-\boldsymbol{x}^i_c \| / R^i$ is the normalized distance from point $\boldsymbol{x}$ to the center $\boldsymbol{x}_c^i$ of the $i$-th local region of radius $R^i$. This function ensures a smooth transition from one inside the region (where $w(z) > 0$) to zero outside, strictly confining the contribution of the local network.

Unlike non-overlapping domain decomposition methods \cite{moseley2023finite}, LENNs do not partition the domain. Instead, the philosophy aligns more closely with modal superposition. The global network captures the fundamental ``mode" of the solution, while the local networks act as corrective ``modes" that are selectively activated near discontinuities to achieve high fidelity. This architecture effectively alleviates spectral bias by allowing different networks to specialize in different frequency components of the solution.

\subsection{Representation of localized discontinuities} \label{DENN}

To equip the local networks within the LENNs framework with the capacity to accurately represent discontinuous behaviors, their inputs are augmented with embedding functions that explicitly encode the geometry of discontinuities. This section details the construction of these embeddings for both strong discontinuities (e.g., cracks) and weak discontinuities (e.g., material interfaces), enabling the neural network to explicitly capture the underlying physical jump conditions.

For strong discontinuities such as cracks, the displacement field itself is discontinuous across the crack surface, as specified by the internal boundary condition on $\Gamma_{d_1}$ in \cref{interface_condition}. To encode this behavior, the input to the local network is augmented with an embedding function $\gamma_s$ constructed from the signed distance function (SDF). The SDF, denoted $\varphi(\boldsymbol{x})$, is defined such that $\varphi(\boldsymbol{x})=0$ on the crack surface, with $\varphi(\boldsymbol{x})>0$ on one side and $\varphi(\boldsymbol{x})<0$ on the other. We construct the embedding function as follows \cite{zhao2025denns}:
\begin{equation}
\gamma_s = w_s(\psi) \cdot \operatorname{sgn}(\varphi),
\end{equation}
where $\operatorname{sgn}(\varphi)$ creates the essential jump across the crack:
\begin{equation}
\operatorname{sgn}(\varphi) =
\begin{cases}
1,  & \text{if } \varphi > 0 \\
-1, & \text{if } \varphi \le 0.
\end{cases} 
\end{equation}
The smooth weight function $w_s(\psi)$ is crucial for ensuring a smooth transition to zero near the crack tip, thereby avoiding the introduction of nonphysical discontinuities. For a center crack, $\psi$ is defined as the product of the distances to two lines passing through the crack tips and perpendicular to the crack line, i.e., $\psi = \psi_1 \cdot \psi_2$. This construct defines a bounded region enveloping the crack. A simple and effective choice for the weight function, adopted in this work, is the squared Rectified Linear Unit:
\begin{equation}
w_s(\psi) = \left( \operatorname{ReLU}(\psi) \right)^2.
\end{equation}
By embedding $\gamma_s$ into the network inputs, a displacement jump is directly induced in the network output across the crack surface:
\begin{equation}
\llbracket \boldsymbol{u} \rrbracket = \llbracket \mathcal{NN}_{\text{local}}(\boldsymbol{x}, \gamma_s) \rrbracket \ne \boldsymbol{0}.
\end{equation}

In contrast, for weak discontinuities such as material interfaces, the displacement field remains continuous across the domain, but its normal derivative exhibits a jump, as mathematically expressed in \cref{interface_condition}. To model this, a distinct embedding function $\gamma_w$ is employed. Following the concept of a ramp function \cite{tseng2023cusp}, we define:
\begin{equation}
\gamma_w = | \phi(\boldsymbol{x}) |,
\end{equation}
where $\phi(\boldsymbol{x})$ is the SDF to the material interface, satisfying $\phi(\boldsymbol{x})=0$ on the interface, with $\phi(\boldsymbol{x})>0$ in $\Omega^+$ and $\phi(\boldsymbol{x})<0$ in $\Omega^-$. This specific construction ensures that $\gamma_w$ is continuous across the interface ($\llbracket \gamma_w \rrbracket = 0$) while possessing a discontinuous normal derivative ($\llbracket \partial \gamma_w/ \partial \boldsymbol{n} \rrbracket \neq 0$). When $\gamma_w$ is passed as an additional input to the local network, the resulting solution $\boldsymbol{u} = \mathcal{NN}_\text{local} (\boldsymbol{x},\gamma_w)$ inherits these properties. Applying the chain rule to the gradient demonstrates that the displacement remains continuous, while its normal derivative exhibits a jump, thereby allowing the network to naturally learn the gradient jump associated with weak discontinuities \cite{zhao2025denns}.

By integrating these geometrically informed embeddings, $\gamma_s$ and $\gamma_w$, into the local networks, the LENNs framework acquires a physics-guided mechanism to accurately represent both strong and weak discontinuities. This approach enhances the representational capacity of PINNs for problems involving localized features, addressing a key limitation of vanilla formulations.

\subsection{Network training procedure}\label{LENN2}

The training of the proposed LENNs is conducted within an energy-based framework. The parameters of the globally coupled neural networks are optimized by minimizing the total potential energy functional of the system, which serves as the physics-informed loss function:
\begin{equation}
\mathcal{L}(\boldsymbol{\theta}_{\text{global}}, {\boldsymbol{\theta}^i_{\text{local}}}) = \Pi(\boldsymbol{u}^*) = \Pi\left( \mathcal{NN}_{\text{global}}(\boldsymbol{x}; \boldsymbol{\theta}_{\text{global}}) + \sum_{i=1}^{n} w_{\text{local}}^{i}(\boldsymbol{x}) \mathcal{NN}_{\text{local}}^{i}\left(\boldsymbol{x}, \gamma^i(\boldsymbol{x}); \boldsymbol{\theta}^i_{\text{local}}\right) \right),
\label{eq:loss function}
\end{equation}
where $\Pi(\cdot)$ is the total potential energy defined in \cref{eq:potential_energy_loss} and $\boldsymbol{u}^*$ is the composite solution defined in \cref{eq:total_solution}. To improve training stability and mitigate potential scale mismatches between the global domain and local subdomains, the spatial coordinates input to the global and each local network are independently normalized to $[-1, 1]$ based on their respective spatial extents. This preprocessing step effectively increases the numerical resolution and sensitivity of the local networks within their designated regions. 

The choice of activation function is critical for ensuring reliable stress evaluation and stable training, as stresses are derived from the derivatives of the displacement field. While ReLU and its variants are popular in deep learning, the discontinuity in their first derivative and the absence of the second derivative renders them unsuitable for accurate stress computation. Consequently, we employ the smooth $\tanh$ function as the activation function throughout all networks. To alleviate the potential gradient vanishing issue associated with $\tanh$ in deep networks, a residual network architecture is adopted. A residual block is defined as \cite{he2016deep}:
\begin{equation}
\boldsymbol{y}^{l+1} = \boldsymbol{y}^{l} + \mathcal{F}(\boldsymbol{y}^{l}; \boldsymbol{\theta}^{l}),
\end{equation}
where $\boldsymbol{y}^{l}$ is the input to the $l$-th block, and $\mathcal{F}$ represents a residual mapping, typically implemented as a sequence of layers (e.g., linear transformation and activation functions) with trainable parameters $\boldsymbol{\theta}^l$. The skip connection facilitates gradient flow during backpropagation, promoting stable training of deeper networks. For the two-dimensional plane stress/strain problems considered in this work, the network is structured to predict the displacement components $\boldsymbol{u} = (u_x, u_y)$ independently (see \cref{LENN}).

While natural boundary conditions are inherently satisfied by the energy formulation, essential boundary conditions must be enforced explicitly. Instead of employing soft penalty terms, which introduce balancing challenges and may not be satisfied exactly, we utilize a hard constraint method. This approach constructs the final network output to satisfy the essential conditions identically, formulated as:
\begin{equation}
\boldsymbol{u}^*(\boldsymbol{x}) = \boldsymbol{B}(\boldsymbol{x}) + \boldsymbol{A}(\boldsymbol{x}) \odot \mathcal{NN}_{\text{raw}}(\boldsymbol{x}),
\end{equation}
where $\mathcal{NN}_{\text{raw}}(\boldsymbol{x})$ is the raw output from the coupled LENNs, $\boldsymbol{B}(\boldsymbol{x})$ is a function constructed to satisfy the non-homogeneous Dirichlet boundary conditions, and $\boldsymbol{A}(\boldsymbol{x})$ is a smooth function that vanishes on the Dirichlet boundary $\Gamma_u$. This construction guarantees that $\boldsymbol{u}^*(\boldsymbol{x}) = \overline{\boldsymbol{u}}(\boldsymbol{x})$ on $\Gamma_u$ exactly. For simple geometries, $\boldsymbol{A}(\boldsymbol{x})$ and $\boldsymbol{B}(\boldsymbol{x})$ can be formulated using distance functions or basic analytical functions \cite{lagaris1998artificial,sukumar2022exact}. 

The evaluation of the energy-based loss functional in \cref{eq:loss function} requires numerical integration over the domain and boundaries. Given the requirement for higher-order continuity in calculating the strain energy, we employ the composite trapezoidal rule for its simplicity and sufficient accuracy in integrating smooth functions approximated by neural networks. The parameters of all networks are optimized simultaneously using gradient-based methods, such as Adam \cite{kingma2017adam}, to minimize the loss function $\mathcal{L}$.

\subsection{Postprocessing: Computation of stress intensity factors}\label{SIF}

\begin{figure}[!hbtp]
    \centering
    \includegraphics[width=0.5\textwidth]{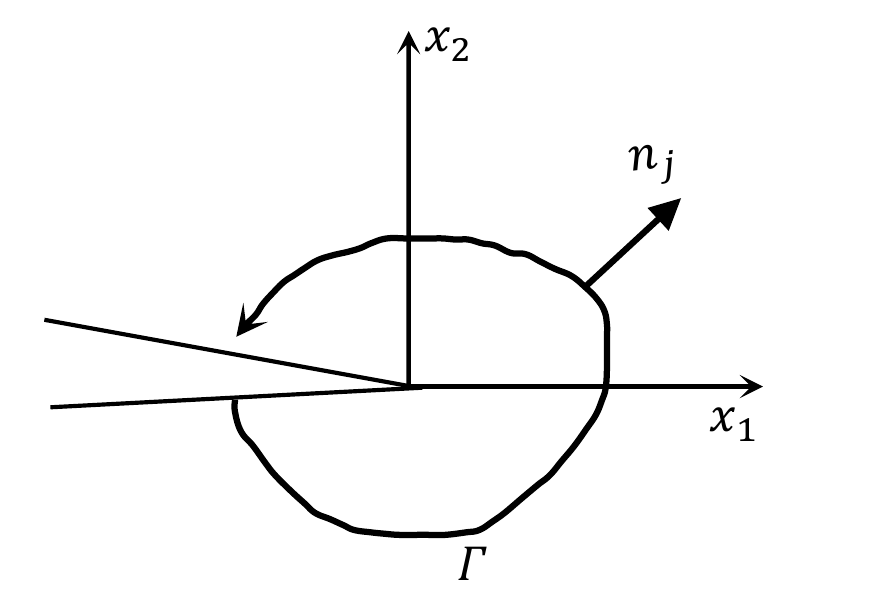}
    \caption{A typical integral path $\Gamma_J$ around a crack tip.}
    \label{J-integral}
\end{figure}

In fracture mechanics, the stress intensity factors (SIFs) are critical indicators that characterize the singularity of the stress field near a crack tip \cite{williams1957stress}. The interaction integral method has been widely adopted for extracting SIFs from numerical solutions. Consider a crack as illustrated in \cref{J-integral}. For an arbitrary smooth contour $\Gamma_J$ surrounding the crack tip, starting from the lower crack surface and traversing counterclockwise to terminate at the upper crack surface, the $J$-integral under quasi-static loading with zero body forces and traction-free crack faces is defined as \cite{rice1968path}:
\begin{equation}
J = \int_{\Gamma_J} \left( W \delta_{1j} - \sigma_{ij} \frac{\partial u_i}{\partial x_1} \right) n_j d\Gamma,
\label{eq:J_integral}
\end{equation}
where $W=\frac{1}{2} \sigma_{ij} \varepsilon_{ij}$ is the strain energy density, $\delta_{1j}$ denotes the Kronecker delta, $n_j$ is the outward unit normal to the contour $\Gamma_J$. The path independence of the $J$-integral holds for homogeneous materials. For bimaterial interface cracks, while properties exhibit discontinuities across the interface, the $J$-integral maintains path independence along directions parallel to the crack tip due to the invariance of material properties in this orientation.

To decouple the mixed-mode SIFs, an auxiliary field $\sigma_{ij}^\text{aux}, \varepsilon_{ij}^\text{aux}, u_i^\text{aux}$ is superposed onto the actual physical field. The $J$-integral for this superimposed state is given by:
\begin{equation}
J^{\text{mix}} = \int_{\Gamma_J} \left[ \frac{1}{2} (\sigma_{kl} + \sigma_{kl}^{\text{aux}})(\varepsilon_{kl} + \varepsilon_{kl}^{\text{aux}}) \delta_{1j} - (\sigma_{ij} + \sigma_{ij}^{\text{aux}}) \frac{\partial (u_i + u_i^{\text{aux}})}{\partial x_1} \right] n_j d\Gamma.
\label{eq:J_superimposed}
\end{equation}
This expression can be decomposed into three components $J^{\text{mix}} = J + J^{\text{aux}} + I$, where $J$ and $J^\text{aux}$ are the $J$-integrals for the actual and auxiliary fields alone, and $I$ is the interaction integral between these two states:
\begin{equation}
I = \int_{\Gamma_J} \left( W^{\text{int}} \delta_{1j} - \sigma_{ij} \frac{\partial u_i^{\text{aux}}}{\partial x_1} - \sigma_{ij}^{\text{aux}} \frac{\partial u_i}{\partial x_1} \right) n_j d\Gamma,
\label{eq:I_integral}
\end{equation}
with $W^{\text{int}} = \frac{1}{2}(\sigma_{ij} \varepsilon_{ij}^{\text{aux}} + \sigma_{ij}^{\text{aux}} \varepsilon_{ij})$ defining the interaction strain energy density.

In conventional mesh-based numerical methods, evaluating the path integral in \cref{eq:I_integral} is challenging due to the singular crack-tip field and discretization limitations. Consequently, the contour integral is typically converted into an equivalent domain integral to improve numerical stability and accuracy. The mesh-free LENNs framework, however, offers an alternative. The high-order continuity of the neural network solutions allows direct evaluation of the path integral, as field derivatives are computed exactly at arbitrary points without interpolation. This bypasses the need for domain transformation, simplifying implementation. It is important to note that this approach remains sensitive to the integration path, as the neural network solution approximates the singularity. Nevertheless, with a proper path selection, accurate results can be achieved, providing a viable and efficient method within this specific framework.

The relationship between the interaction integral and SIFs is established through energy considerations. For linear elastic fracture in homogeneous media, the $J$-integral equals the energy release rate $G$, which is related to the mixed-mode SIFs $K_\text{I}$ and $K_\text{II}$ by:
\begin{equation}
J = G = (K_\text{I}^2 + K_\text{II}^2)/{E^*}, 
\end{equation}
where $E^* = E$ for plane stress and $E^* = E/(1-\nu^2)$ for plane strain, and $E$ and $\nu$ represents the Young's modulus and Poisson's ratio, respectively. Selecting the crack-tip asymptotic field as the auxiliary field, the interaction integral $I$ is linked to the SIFs through the following relation:\begin{equation}
I = \frac{2}{E^*} \left(K_\text{I} K_\text{I}^{\text{aux}} + K_\text{II} K_\text{II}^{\text{aux}}\right).
\label{eq:I_sif_relation}
\end{equation}

The individual SIF components can be extracted by strategically selecting the auxiliary SIFs. Setting $(K_\text{I}^\text{aux}, K_\text{II}^\text{aux}) = (1, 0)$ yields the interaction integral $I^{(1,0)}$, and setting $(K_\text{I}^{\text{aux}}, K_\text{II}^{\text{aux}}) = (0, 1)$ yields $I^{(0,1)}$. The mode-I and mode-II SIFs are then computed as:
\begin{equation}
K_\text{I} = \frac{E^*}{2} I^{(1,0)}, \quad K_\text{II} = \frac{E^*}{2} I^{(0,1)}.
\label{eq:SIFs_final}
\end{equation}
This formulation provides an accurate and computationally efficient means for SIF evaluation within the proposed LENNs framework.

\section{Numerical results}\label{Results}

This section presents a comprehensive evaluation of the proposed LENNs framework through a series of benchmark problems in solid mechanics. The primary objectives are twofold: to demonstrate the capacity of LENNs in accurately resolving both localized strong and weak discontinuities, and to conduct a comparative analysis against established methods, including the standard DEM, DENNs, and CENN. To ensure a fair and consistent comparison, a standardized neural network configuration is employed across all experiments unless otherwise specified. The architecture and hyperparameters, determined through preliminary studies to offer a robust balance between performance and efficiency, are summarized as follows. The global network is configured with 4 hidden layers and 25 neurons per layer, while the local network employs a deeper architecture of 6 hidden layers with 25 neurons per layer to enhance its capacity for modeling complex local features. All models are trained with an initial learning rate of 0.008, which is reduced by a factor of 0.5 every 5000 epochs. An early stopping criterion is applied if the loss does not improve over 2000 consecutive epochs. The implementation is carried out using the PyTorch library \cite{paszke2019pytorch}, with computations performed on NVIDIA GeForce RTX 4060 Ti GPU.



\subsection{ A short center crack in a tensile plate} \label{example111}

\begin{figure}[!hbtp]
    \centering
    \includegraphics[width=1.0\textwidth]{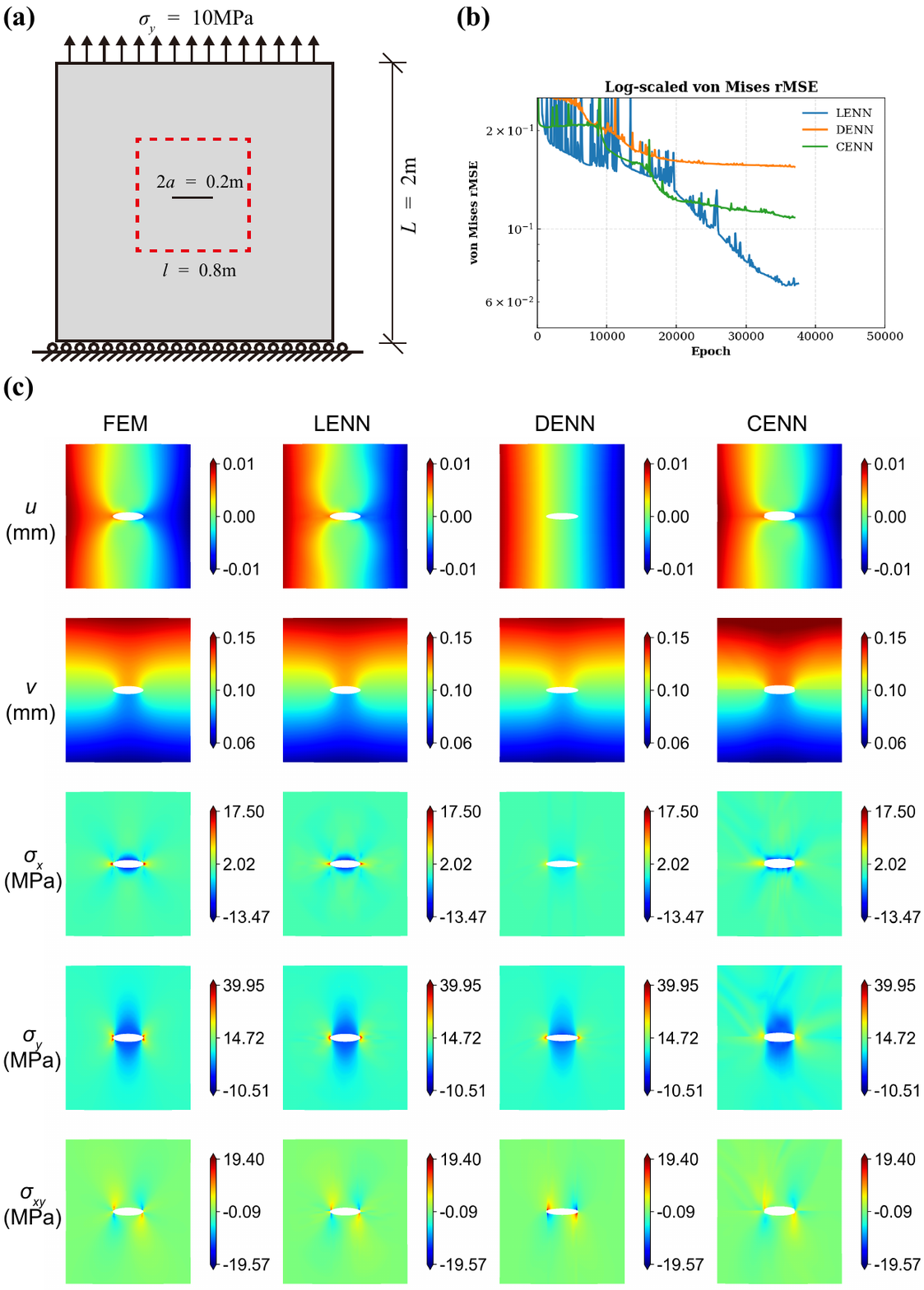}
    \caption{Short center crack under tensile loading: 
    (a) Geometry of the symmetric part selected for computation; 
    (b) Evolution of rRMSE of von Mises stress over training epochs for LENN, DENN, CENN within the selected subregion; 
    (c) Comparison of displacement and stress components of LENN, DENN, CENN and FEM in the selected subregion.
    }
    \label{example1}
\end{figure}

Given the significance of cracks in structural failure analysis, we consider a short crack problem to assess the ability of LENNs to handle localized discontinuities and stress singularities, as illustrated in \cref{example1}(a). This example is designed to examine whether the method can accurately capture displacement jumps across the crack surface and stress concentration near the crack tips.
The computational domain is a plate defined on \( [-1, 1] \times [-1, 1] \), containing a short crack centered on \( (0, 0) \) with a half-length of 0.1. The Young's modulus is set to be \( 10^5\,\text{MPa} \), and the Poisson's ratio is set to be 0.3. A plane-stress condition is assumed and a tensile load \( \sigma_y = 10\,\text{MPa} \) is applied to the upper boundary.

We employ two different network architectures that are mentioned in \cref{LENN1} to address this problem. One is referred to as the global network, which maps \( (x, y) \) to \( (u, v) \); the other is called the local network, which incorporates an additional coordinate \( \gamma_s \) in its input (resulting in three input dimensions) to capture discontinuities more accurately. The additional coordinate is specifically represented as:
\begin{equation}
\gamma_s = \text{ReLU}(-\psi_1 \psi_2) \cdot \text{sgn}(\varphi)
\end{equation}
where \( \psi_1 \) and \( \psi_2 \) denote the distances from a point within the domain to two lines that originate at the crack tips and extend in directions normal to the crack surface, while \( \varphi \) represents the distance from any point within the domain to the crack surface. We designate the local domain as a circular region with a radius equal to four times the crack half-length. A total of \( 300 \times 300 \) grid points are employed within the domain for training. It should be mentioned that all inputs of the neural networks were normalized to the range of \([-1, 1]\).

To evaluate the efficiency and accuracy of LENNs, we perform comparative tests with CENNs \cite{wang2022cenn} and DENNs \cite{zhao2025denns}. For comparison, CENNs and DENNs are implemented with network architectures of similar scale and hyperparameters, including depth, number of neurons, activation functions, and learning rate schedules. The number of collocation points of these three methods is the same. In all approaches, hard constraints are applied to enforce the essential boundary conditions.

\cref{table_example1} summarizes the network configurations and training time, thereby providing a quantitative basis for performance comparison. To highlight the differences in predictive performance near the crack tip, the comparison is confined to the red dashed subregion
in \cref{example1}(a). \cref{example1}(c) presents the displacement and stress solutions obtained by the three methods and the results indicate that DENN performs inaccurate predictions in the displacement field, particularly for the secondary component of the displacement \(u\);
Meanwhile, CENN produces inaccurate predictions of the stress field and cannot capture the stress concentration near the crack tips. In contrast, leveraging the fine-resolution capability of its local network, LENN accurately identifies the crack and effectively learns both displacement and stress fields in the crack tip region, with results that closely match the results obtained from COMSOL finite element simulations.
\begin{table}[!hbtp]
    \begin{center}
    \caption{\label{table_example1} Comparative analysis of hyperparameters across CENNs, DENNs, LENNs, and FEM for cracks.}     
    \begin{tabular}{cccccc}
        \toprule
        & & CENNs & DENN & LENNs & FEM \\
        \midrule
        \multirow{7}{*}{Hyperparameters} 
         & subdomains & 2 & / & / & / \\
         & trainable parameters & 10804 & 8770 & 8254 & / \\
         & internal points  & 90000 & 90000 & 90000 & 65030 (nodes) \\
         & boundary points & 3000 & 256 & 256 & / \\
         & interface points & 2000 & / & /  & / \\
         & training epochs & 37500 & 37500 & 37500 & / \\
         & time (seconds/100 epochs) & 5.09 & 4.24 & 4.95 & / \\
        \bottomrule
    \end{tabular}  
    \end{center}
\end{table}

Furthermore, \cref{example1}(b) also illustrates the evolution of the relative root mean square error (rRMSE) of the von Mises stress within the same selected region, compared to finite element solutions, over training epochs for all three methods. The rRMSE of a physical field \( \Phi \) is defined as:
\begin{equation}
\text{rRMSE}(\Phi) = \sqrt{
\frac{
\int_{\Omega} \left\| \Phi_{\mathcal{NN}} - \Phi_{\text{FEM}} \right\|^2 \, d\Omega
}{
\int_{\Omega} \left\| \Phi_{\text{FEM}} \right\|^2 \, d\Omega
}
}
\end{equation}
These curves clearly demonstrate that LENN consistently achieves lower stress prediction errors, indicating its superior accuracy and robustness in handling localized features such as stress singularities and discontinuities.

\begin{figure}[!hbtp]
    \centering
    \includegraphics[width=1.0\textwidth]{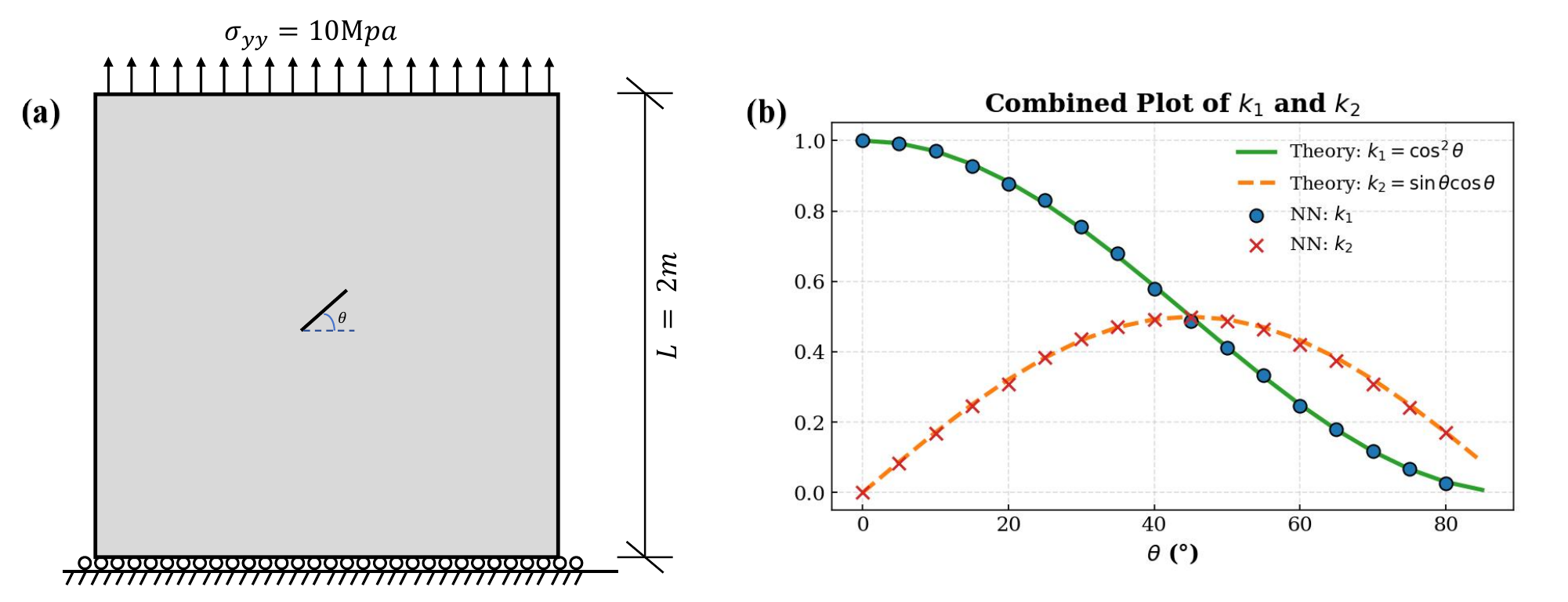}
    \caption{Short center crack under tensile loading: 
    (a) crack angle \(\theta\) varies, ranging from \(0^\circ\) to \(90^\circ\); 
    (b) Stress intensity factor \( K_1 \) and \( K_2 \) versus angle variation curve: the blue dot represents the predicted value of \( K_1 \), the red cross represents the predicted value of \( K_2 \), the green solid line represents the theoretical value of \( K_1 \), and the orange dashed line represents the theoretical value of \( K_2 \).
    }
    \label{example1-2}
\end{figure}

\begin{table}[htbp]
    \begin{center}
    \caption{$K_I$ values and relative errors computed from the J-integral under different integration radius}
    \label{table2_example1}
    \begin{tabular}{cccc}
        \toprule
        Method & Integration radius $r$ & $K_I$ & Relative Error (\%) \\
        \midrule
        \multirow{4}{*}{LENN} 
         & 0.02  & 178.48 & 0.7 \\
         & 0.04  & 177.11 & 0.1 \\
         & 0.05  & 178.83 & 0.9 \\
         & 0.075 & 178.13 & 0.5 \\
        \midrule
        \multirow{1}{*}{DENN} 
         & 0.075 & 166.98 & 5.79 \\
        \bottomrule
    \end{tabular}
    \end{center}
\end{table}

To further explore the performance of LENNs and quantify how effectively the method captures crack tip features, stress intensity factors (SIFs) are extracted, which play an important role in linear elastic fracture mechanics. Specifically, the SIF characterizes how stress behaves in the vicinity of a crack tip, where a mathematical singularity arises in the classical continuum framework. In this study, we keep the crack length the same but vary the crack angle from \( 0^\circ \) to \( 90^\circ \), and use the \( J \)-integral\cite{yau1980mixed} to calculate the SIF and compare them with the theoretical values in the reference \cite{moes1999finite}. The accuracy of the \( J \)-integral implementation is first verified at \( 0^\circ \), showing excellent agreement with the theoretical solution in \cite{moes1999finite}, as summarized in \cref{table2_example1}. We then compute the SIFs for crack angles ranging from \( 0^\circ \) to \( 90^\circ \), with an interval of \( 5^\circ \). \cref{example1-2}(b) presents the computed SIFs as a function of the crack angle. The results remain highly consistent with the reference values \cite{moes1999finite}.

Overall, for short-crack problems characterized by strong discontinuities and stress singularities, LENN not only achieves more accurate displacement and stress fields but also enables reliable extraction of SIFs. These findings demonstrate the robustness and precision of LENN in capturing complex fracture mechanics behavior.

\subsection{A bi-material plate with partial debonding}

\begin{figure}[!hbtp]
    \centering
    \includegraphics[width=0.6\textwidth]{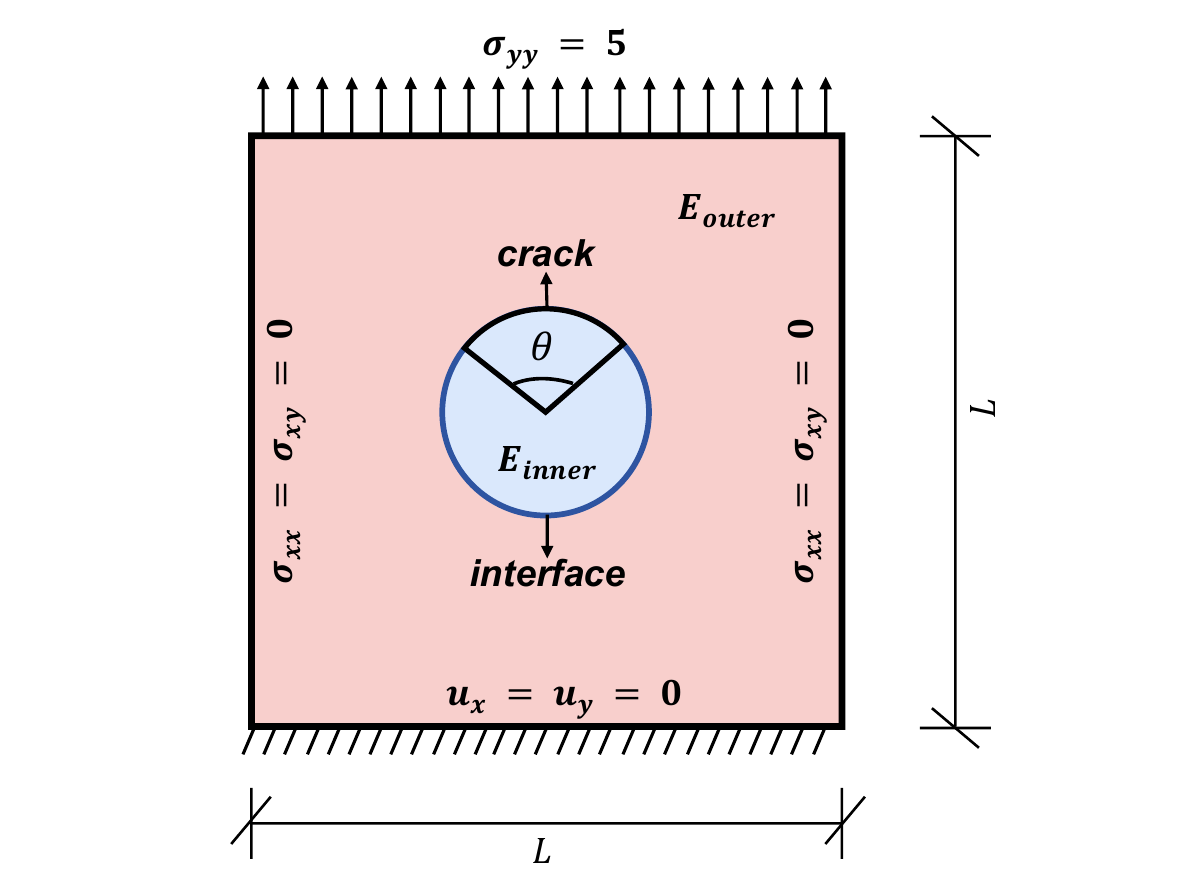}
    \caption{Geometry of the symmetric part selected for computation.
    }
    \label{example2-1}
\end{figure}

In this example, to evaluate the capability of LENNs in handling weak discontinuities and capturing stress jumps across material interfaces with partial bonding failure, we introduce a representative problem involving both material heterogeneity and interfacial debonding. We consider a two-dimensional bi-material structure where the outer rectangular region and the inner circular region possess different material properties. Specifically, the rectangular domain is \( [0,1] \times [0,1] \), while the circular inclusion has a center at \( (0.5, 0.5) \) with a radius of \( 0.15\,\text{m} \). The Young's modulus in the circular region is \( 10^4\,\text{MPa} \), whereas that of the outer domain is \( 10^3\,\text{MPa} \); both regions share the same Poisson's ratio of 0.3. We impose a tensile stress of \( 5\,\text{MPa} \) on the top boundary, while the bottom boundary is fixed, that is, \( (u,v) = (0,0) \). To model interfacial debonding, an interfacial crack is introduced along the circular boundary of the inclusion as shown in \cref{example2-1}, with the crack angle set to \( \theta = 120^\circ \).

Similarly to the previous strong-discontinuity examples, we train two networks---the global network and the local network. We adopt a hard-constraint approach for displacement boundary conditions, as described in \cref{example111}, which can be expressed as:
\begin{equation}
u_{\text{out}} = u_{\mathcal{NN}} \cdot y, \quad 
v_{\text{out}} = v_{\mathcal{NN}} \cdot y
\end{equation}
The global network adopts the same architecture as in previous examples, while the local network is modified to embed two additional input coordinates, one capturing the strong discontinuity caused by the crack, and the other modeling the weak discontinuity at the material interface---thereby enhancing the network's capacity to resolve crack-tip features. The embedding functions can be expressed as:
\begin{align}
\boldsymbol{\gamma} &= [\gamma_w, \gamma_s] \label{eq:gamma_def} \\
\gamma_w &= \left| \varphi \right|, \quad 
\gamma_s = \text{ReLU}(-\psi_1 \psi_2) \cdot \text{sgn}(\varphi') \label{eq:embedding_def}
\end{align}
where \( \varphi \) is defined as:
\begin{equation}
\varphi(x, y) = (x - x_0)^2 + (y - y_0)^2 - r_0^2
\end{equation}
and \( \varphi' \) is defined based on \( \varphi \), sharing the same form of the function but restricted to a specified arc segment of the circular interface, corresponding to the interfacial crack, \( \psi_1 \) and \( \psi_2 \) denote the perpendicular distances from any point in the domain to the two endpoints of the arc segment along lines normal to the circular interface. The final network output is multiplied by a window function, restricting it to the vicinity of the circular domain with radius \( r_{\text{local}} = 1.5 \, r_{\text{circle}} \), to ensure the local network only affects its corresponding region. In this example, we utilize a total of \( 300 \times 300 \) sampling points over the entire domain for both training and numerical integration.

\begin{figure}[!hbtp]
    \centering
    \includegraphics[width=1.0\textwidth]{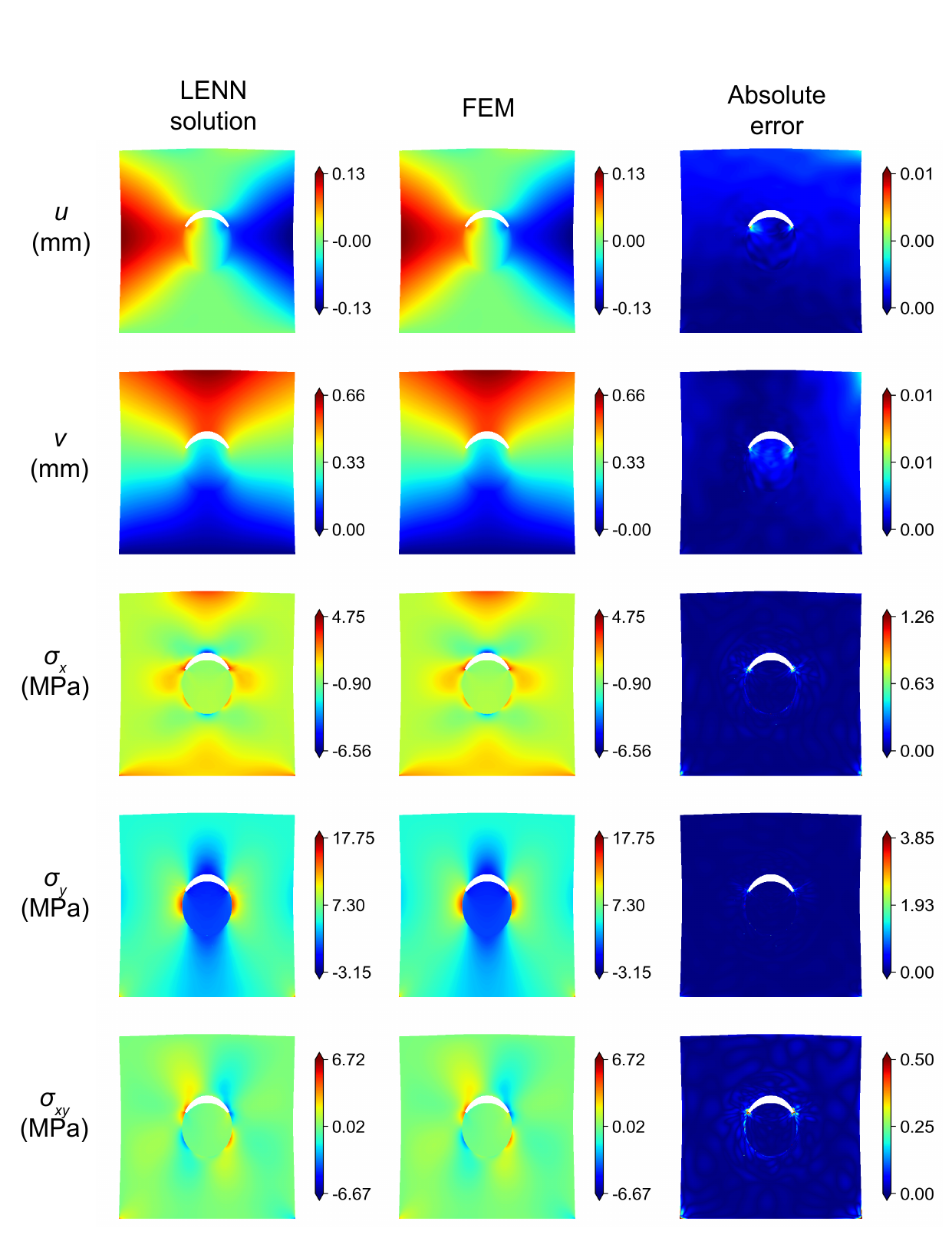}
    \caption{Comparison of displacement and stress fields predicted by LENNs with COMSOL results for interfacial debonding, showing accurate capture of interface discontinuity and stress concentration.
    }
    \label{example2-2}
\end{figure}

To evaluate the performance of LENNs in modeling interfacial debonding, we compared the predicted displacement and stress fields against finite element results obtained from COMSOL. \cref{example2-2} shows the distributions of the displacement and stress components, respectively. As observed, LENNs capture the discontinuity across the material interface and accurately reproduce the stress concentration near the debonding region. The displacement field exhibits a smooth transition across the bonded interface and a distinct jump along the debonding segment, which is consistent with the expected physical behavior.

Subsequently, we further vary other properties over a range of values and measure the change in performance, this is carried out for the stiffness ratio between the inner and outer materials from 0.1 to 10, and the interfacial crack angle from \( 60^\circ \) to \( 180^\circ \). To quantify accuracy, we compute the relative root mean square error (rRMSE) of the stress field under different configurations. \cref{table1_example2} lists the rRMSE values for several test cases with varying material property contrasts and crack orientations. As shown, the accuracy of LENN does not depend significantly on the stiffness ratio between the inner and outer materials or notable changes in the interfacial crack angle. The method maintains low error levels across all cases, demonstrating strong generalization and robustness for problems involving both material heterogeneity and interfacial cracks.

\begin{table}[htbp]
    \begin{center}
    \caption{Displacement rRMSE under different Young's modulus ratios and interface angles}
    \label{table1_example2}
    \begin{tabular}{ccc}
        \toprule
        Young's modulus ratio ($E_{\text{in}}/E_{\text{out}}$) & Interface angle  & Displacement rRMSE \\
        \midrule
        \multirow{3}{*}{0.1} 
            & $60^\circ$  & $3.2 \times 10^{-3}$ \\
            & $90^\circ$  & $3.8 \times 10^{-3}$ \\
            & $120^\circ$ & $5.5 \times 10^{-3}$ \\
        \midrule
        \multirow{3}{*}{2.5} 
            & $60^\circ$  & $2.4 \times 10^{-3}$ \\
            & $90^\circ$  & $3.7 \times 10^{-3}$ \\
            & $120^\circ$ & $5.5 \times 10^{-3}$ \\
        \midrule
        \multirow{3}{*}{10} 
            & $60^\circ$  & $3.6 \times 10^{-3}$ \\
            & $90^\circ$  & $8.6 \times 10^{-3}$ \\
            & $120^\circ$ & $5.2 \times 10^{-3}$ \\
        \bottomrule
    \end{tabular}
    \end{center}
\end{table}

\subsection{Quasi-static propagation of edge cracks} 

Building on the ability of LENNs to accurately compute stress intensity factors, as demonstrated in \cref{example111}, we now turn to a more advanced and practically relevant problem: the prediction of crack propagation paths. In this example, we assess the capability of LENNs in simulating crack propagation driven by displacement loading.

\begin{figure}[!hbtp]
    \centering
    \includegraphics[width=1.0\textwidth]{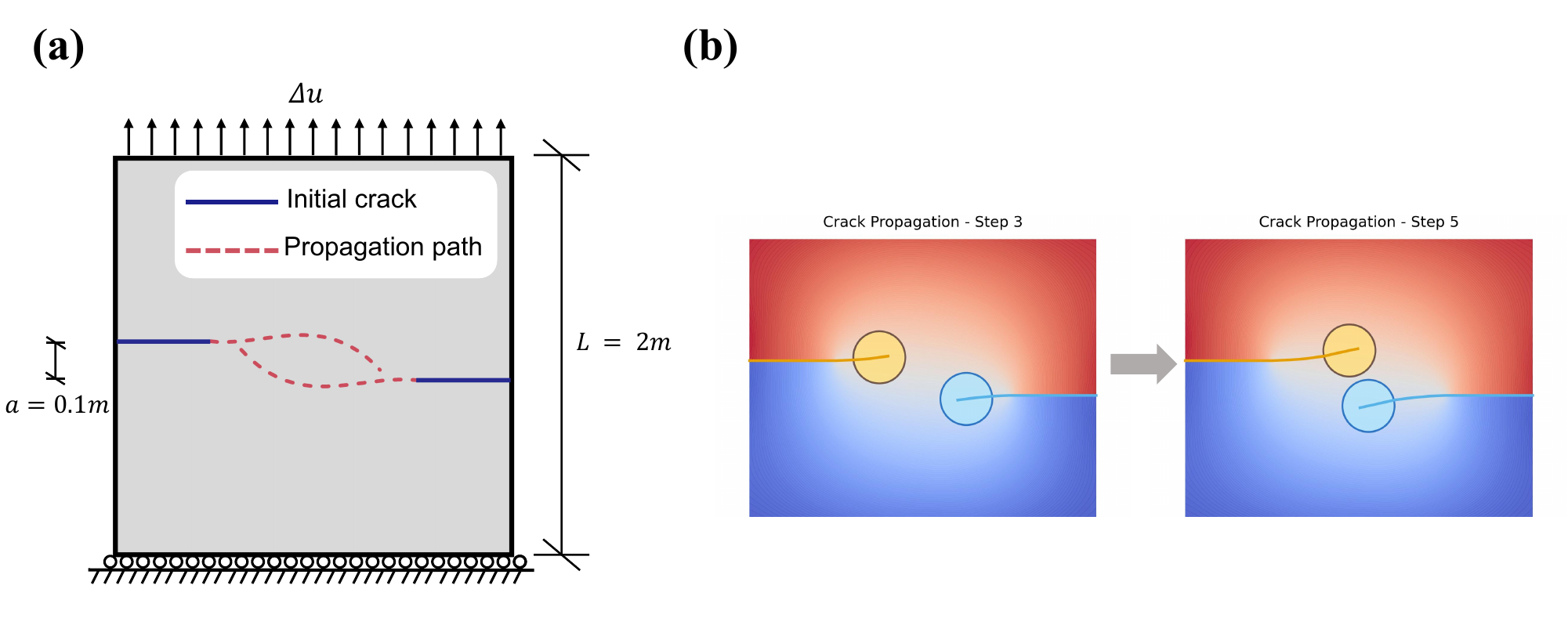}
    \caption{Edge crack growth under displacement loading: 
        (a) Geometry of the symmetric part selected for computation; 
        (b) Schematic of local network regions moving with crack propagation; 
    }
    \label{example3-1}
\end{figure}

The computational domain contains two edge cracks, one located on the left boundary and the other on the right boundary of the specimen, as shown in \cref{example3-1}(a). A displacement boundary condition is applied on the top edge, while the bottom edge is fixed in the vertical direction. To prevent rigid body motion, an additional constraint is imposed at one point on the bottom boundary. The elastic properties of the plate are defined by \( E = 2 \times 10^5\, \text{MPa} \) and \( \nu = 0.3 \).

Crack growth is modeled as a quasi-static process to ensure stable propagation throughout the simulation. At each step, the maximum circumferential stress criterion \cite{Erdogan1963Crack} is employed to determine the crack extension direction. Further theoretical and numerical details are provided in \cref{AppendixB}. A propagation simulation is conducted with a fixed incremental crack length of 0.1. Since multiple crack segments arise during the propagation process, to accurately represent the discontinuities arising from multiple cracks, the embedding function must be appropriately adjusted. For simplicity, we illustrate the construction of the embedding using the propagation path of one representative crack segment. The embedding function is constructed as:
\begin{equation}
\gamma = \left[ \left( \text{ReLU}^2(-\psi_2) - 1 \right) \cdot \text{Heaviside}(\psi_3) + 1 \right] \cdot \text{sgn}(\varphi)
\end{equation}
where $\psi_2(x)$ denotes the signed distance from any point to the line segment that is perpendicular to the final crack segment and passes through its endpoint; $\psi_3(x)$ represents the signed distance to a line perpendicular to the final crack and passing through its initial point. The function $\varphi(x)$ is defined as the signed distance from the point to the current entire crack path, including both the initial cracks and all segments generated during propagation. It takes positive values above the crack and negative values below. The Heaviside function is defined as:
\begin{equation}
\text{Heaviside}(x) =
\begin{cases}
1, & x > 0 \\
0, & x \leq 0
\end{cases}
\end{equation}
After each propagation step, the embedding function is dynamically updated to reflect the new crack geometry and ensure an accurate representation of the discontinuity during subsequent training.
\begin{figure}[!hbtp]
    \centering
    \includegraphics[width=1.0\textwidth]{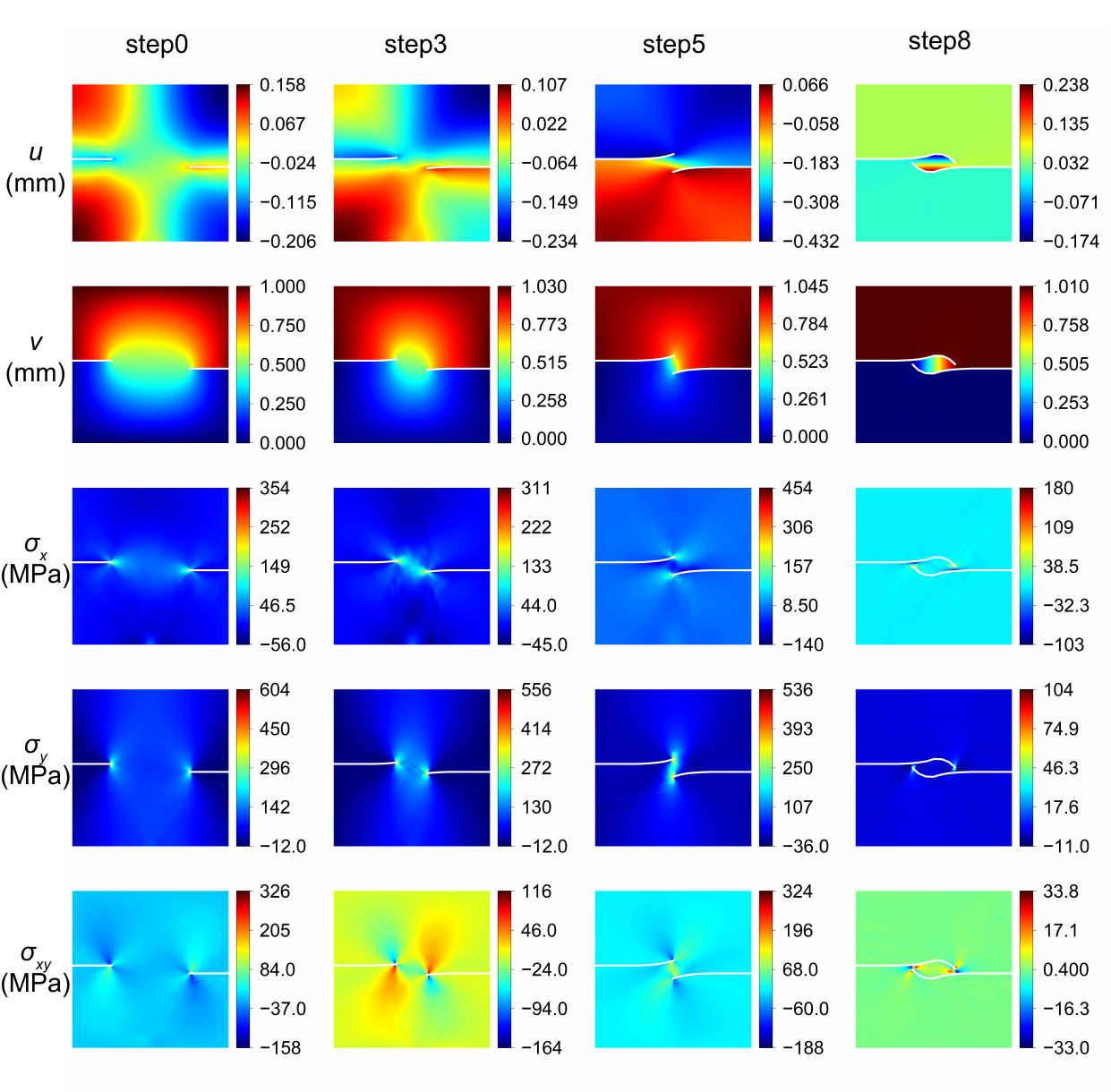}
    \caption{Crack propagation results predicted by LENN, including crack paths, displacement and stress distributions, showcasing its fidelity in modeling dynamic fracture behavior. 
    }
    \label{example3-2}
\end{figure}


For the network structure, unlike previous cases, in this example two distinct local networks are used to describe the corresponding edge cracks. The whole network are trained with \( 350 \times 350 \) collocation  points regularly spaced throughout the domain. Following the previous setup, all inputs of the neural networks are normalized to the range of \([-1, 1]\). Due to the similarity of the physical configuration across propagation steps, the network trained in the first step is reused as the initial model for all subsequent steps. This transfer learning\cite{goswami2020transfer} strategy not only reduces the total training time required for later steps by providing a well-preconditioned initialization already adapted to the characteristics of the fracture problem but also ensures consistent initialization throughout the evolving crack geometry. Meanwhile, the effective regions of the local networks move along with the crack propagation, with each region centered at the midpoint of the corresponding newly extended crack segment, as shown in \cref{example3-1}(b).

As shown in \cref{example3-2}, the predicted crack path with an incremental crack length of 0.1 is presented, together with the associated stress and displacement distributions at each step. Detailed results are reported for this case. These findings demonstrate that LENNs are capable of capturing not only static fracture fields but also dynamic crack growth processes, providing reliable predictions of propagation direction and quantitative accuracy.

\section{Discussion and conclusion} \label{conclusion}

In this work, we proposed a Locally Enhanced Neural Networks framework for solving solid mechanics problems with localized discontinuous features such as cracks and material interfaces. The method is based on a multiscale PINN architecture, which combines a global neural network that models the smooth global solution and local network is designed to capture high-gradient or discontinuous features. This separation reduces the interference between global and local behaviors during training. The two components are fused through spatial window functions to maintain consistency across the solution domain. In addition, the local network leverages embedding strategies to represent both strong and weak discontinuities effectively, allowing for sharp behaviors in the solution field.

Several numerical experiments have been conducted to evaluate the performance of LENNs. The results demonstrate that LENNs achieve high accuracy in predicting both displacement and stress fields, particularly near discontinuities, where it consistently outperforms existing PINN-based methods such as CENNs and DENNs. Through simulations involving cracks at different orientations and interface debonding with varying material properties, LENNs exhibit better generalization capability across diverse discontinuity configurations. During the experiments, we also observed that the use of local enhancement leads to a more stable training process, especially in cases involving high-gradient or singular regions. Furthermore, we explore transfer learning into the LENN framework for crack propagation scenarios, and find that it not only accelerates convergence in subsequent propagation steps but also ensures consistency in crack growth process.

Despite the success exhibited by the LENN framework, several aspects warrant further investigation and discussion. First, during the investigation of bi-material circular inclusion problems, we observed that when the Young's modulus of the inner region is significantly higher than that of the outer region, stress oscillations appear within the circular domain. This issue persists regardless of whether the radius of the inclusion is increased or decreased. We hypothesize that this may be caused by overfitting of the local network within the circular domain. Two potential strategies could be considered to address this: (1) improving the formulation of weak discontinuity embedding function to better reflect the physical behavior at the material interface; (2) altering the training pattern, i.e., instead of training the global and local networks simultaneously, we could first train the global network to capture the coarse-scale response, and subsequently train the local network to resolve fine-scale features.

Second, due to the use of energy-based loss formulations such as the Deep Energy Method (DEM), the numerical accuracy of integration plays a critical role in overall performance. In the present work, we still adopt a uniformly spaced composite trapezoidal integration scheme for numerical computation. However, in cases involving steep stress gradients or singularities, insufficient integration resolution may lead to underestimation of energy and degradation of training accuracy. Employing adaptive integration schemes or higher-order quadrature rules could help mitigate this issue. Previous studies\cite{wang2025kolmogorov} have also shown that employing triangular numerical integration can significantly enhance the predictive accuracy of physics-informed neural networks.

Third, in the current framework, the definition of local subdomains is manually prescribed based on prior knowledge of the problem, such as the expected location of discontinuities or stress concentrations. While this strategy has proven effective in practice, it limits the general applicability of the method and introduces dependence on user expertise. To improve adaptability and accuracy, it would be beneficial to develop an automated mechanism capable of dynamically identifying and adjusting the local enrichment regions according to the evolving features of the solution. Moreover, although we have experimented with various window functions---such as cubic B-splines, sigmoid functions---and found that the choice has limited influence on the overall performance, the possibility remains that alternative formulations of window functions could provide improved numerical stability or learning efficiency, particularly in problems involving sharp gradients or multiscale interactions.

Overall, the proposed LENN framework provides a promising foundation for advancing physics-informed neural networks in the modeling of fracture and localized discontinuities within complex heterogeneous materials. Its modular and localized design supports scalability and parallel implementation, paving the way for efficient large-scale and multiscale simulations. Potential extensions include integration with multiphysics formulations---such as thermo-mechanical fracture---and adaptive allocation of local networks to reduce computational redundancy and improve accuracy in critical regions.

\section*{Data availability}
The implementation of our method, along with all relevant code, will be released on GitHub after the paper is officially published. The repository link will be provided in the final version.

\section*{Acknowledgments}
This work has been supported by the National Natural Science Foundation of China (Grant No.~12272277) and the Guangdong Basic and Applied Basic Research Foundation (Grant No. 2023A1515030161).

\newpage
\appendix

\section{Maximum circumferential stress criterion}\label{AppendixB}

\renewcommand{\theequation}{A.\arabic{equation}}
\setcounter{equation}{0}
\renewcommand{\thefigure}{A.\arabic{figure}}
\setcounter{figure}{0}
\renewcommand{\thetable}{A.\arabic{table}}
\setcounter{table}{0}

To predict the direction of crack growth, we adopt the maximum circumferential stress criterion, which assumes that the crack extends in the direction where the hoop stress \( \sigma_{\theta\theta} \) reaches its maximum near the crack tip. Under plane strain conditions, the expression for \( \sigma_{\theta\theta} \) in polar coordinates around the crack tip is:
\begin{equation}
\sigma_{\theta\theta}(r, \theta) = \frac{1}{\sqrt{2\pi r}} 
\left[
K_I \cos\left(\frac{\theta}{2}\right)\left(1 - \sin\left(\frac{\theta}{2}\right) \sin\left(\frac{3\theta}{2}\right)\right)
- K_{II} \sin\left(\frac{\theta}{2}\right)\left(2 + \cos\left(\frac{\theta}{2}\right) \cos\left(\frac{3\theta}{2}\right)\right)
\right]
\end{equation}
To determine the direction \( \theta_c \) in which the crack will propagate, we take the derivative of \( \sigma_{\theta\theta} \) with respect to \( \theta \), and set it to zero:
\begin{equation}
\frac{\partial \sigma_{\theta\theta}}{\partial \theta} = 0
\end{equation}
This yields the condition:
\begin{equation}
K_I \sin \theta_c + K_{II} (3 \cos \theta_c - 1) = 0
\end{equation}
Solving this equation gives:
\begin{equation}
\theta_c = \arccos \left( 
\frac{3K_{II}^2 \pm \sqrt{K_I^4 + 8K_I^2 K_{II}^2}}{K_I^2 + 9K_{II}^2}
\right)
\end{equation}

%
\bibliographystyle{elsarticle-num}
\bibliography{scopus}
\end{document}